\documentclass[twocolumn, journal]{IEEEtran}

\IEEEoverridecommandlockouts

\usepackage{float}
\usepackage{stfloats}
\usepackage{mathrsfs}
\usepackage{diagbox}
\usepackage{bm}
\usepackage{color}
\usepackage{graphicx}
\usepackage{amsmath}
\usepackage{amssymb}
\usepackage{algorithm}
\usepackage{algorithmic}
\usepackage{ulem}
\usepackage{lineno}
\usepackage{lettrine}
\usepackage{subfigure}
\usepackage{epstopdf}
\usepackage{stfloats}
\usepackage{color}
\usepackage{graphicx}
\usepackage{amsmath}
\usepackage{amssymb}
\usepackage{algorithm}
\usepackage{algorithmic}
\usepackage{amsmath}
\usepackage{multirow}
\usepackage{booktabs}
\usepackage{array}
\usepackage{amsthm}
\usepackage{stfloats}
\usepackage{caption}
\usepackage{subfigure}
\usepackage{bm}
\usepackage{booktabs}
\usepackage{setspace}
\usepackage{makecell}
\usepackage{cases}

\newcommand{\be}{\begin{equation}}
	\newcommand{\ee}{\end{equation}}
\newcommand{\bea}{\begin{eqnarray}}
	\newcommand{\eea}{\end{eqnarray}}
\newcommand{\ba}{\begin{array}}
	\newcommand{\ea}{\end{array}}

\renewcommand{\thesubsubsection}{\thesubsection.\arabic{subsubsection}}

\allowdisplaybreaks[4]

\title{UD-MAC: Delay Tolerant Multiple Access Control Protocol for Unmanned Aerial Vehicle Networks
	\thanks{ Yingying Zou, Zhiqing Wei, Yanpeng Cui and Zhiyong Feng, are with Key Laboratory of Universal Wireless Communications, Ministry of Education, School of Information and Communication Engineering, Beijing University of Posts and Telecommunications (BUPT), Beijing 100876, China (e-mail: zouyingying@bupt.edu.cn, weizhiqing@bupt.edu.cn, cuiyanpeng94@126.com, fengzy@bupt.edu.cn}
	\thanks{ Xinyi Liu is with Ali Travel Network Technology (Beijing) Co., Ltd., Beijing 100102, China (e-mail: xinzhi.lxy@alibaba-inc.com).}}

\author{Yingying Zou, Zhiqing Wei, Yanpeng Cui, Xinyi Liu, and Zhiyong Feng
}
\begin{document}
\maketitle
\thispagestyle{empty}
\begin{abstract}
In unmanned aerial vehicle (UAV) networks, high-capacity data transmission is of utmost importance for applications such as intelligent transportation, smart cities, and forest monitoring, which rely on the mobility of UAVs to collect and transmit large amount of data, including video and image data. Due to the short flight time of UAVs, the network capacity will be reduced when they return to the ground unit for charging. Hence, we suggest that UAVs can apply a store-carry-and-forward (SCF) transmission mode to carry packets on their way back to the ground unit for improving network throughput. In this paper, we propose a novel protocol, named UAV delay-tolerant multiple access control (UD-MAC), which can support different transmission modes in UAV networks. We set a higher priority for SCF transmission and analyze the probability of being in SCF mode to derive network throughput. The simulation results show that the network throughput of UD-MAC is improved by 57\% to 83\% compared to VeMAC.	
\end{abstract}
	
	
\section{Introduction}
\label{sec:introduction}
With the deep integration of communication and sensing technology, there is an increasing demand for data collection and transmission. Because it is difficult to deploy sensors in complex scenarios such as forests and oceans, unmanned aerial vehicles (UAVs) have become the most popular platforms for sensing devices due to their mobility \cite{intro_1}. UAV wireless sensor network (UWSN) can not only overcome the limitations of data collection in complex geographical environments \cite{intro_2,intro_3,intro_4,intro_5,new_7}, but also be deployed in hazardous environments for tasks to reduce human injury \cite{new_1}\cite{intro_6}. Therefore, we are at the dawn of the era of ubiquitous aerial communication and sensing, in which UAVs with onboard sensors and compute units are connected seamlessly to enable profound progress in military and civilian applications \cite{intro_7}\cite{new_5}. The large amount of data, such as video and images, required by the UWSN applications described above presents a challenge for high-capacity data transmission in UAV networks.

In UWSN, the delay tolerant network (DTN) can meet the needs of intermittent connections due to the high mobility and sparsity of UWSN \cite{intro_71}. To meet the challenges of high-capacity data transmission, many scholars effectively improve throughput through path planning \cite{new_3,new_4,new_6,intro_8,intro_9,intro_10}, routing protocols \cite{intro_11,intro_12,intro_13}, and multiple access control (MAC) schemes\cite{new_2}. The path planning requires a trade-off between network throughput and energy consumption since the flight time of UAVs is limited by the onboard energy capacity \cite{intro_14}. Therefore, it cannot fundamentally improve network throughput. Various routing schemes have been proposed for DTN aiming at increasing the message delivery probability \cite{intro_72}. But these schemes pay little attention to high-capacity transmission. They also require reliable and stable communication link support, which is hard to be satisfied in UAV networks with highly dynamic topologies \cite{intro_11}. Unlike the routing protocols that focus on end-to-end transmission, MAC schemes focus on point-to-point transmission, which increases the probability of successful transmission by solving the collision problem during transmission, thereby radically increasing network throughput. Therefore, designing an efficient MAC scheme is the priority of throughput enhancement.

Motivated by the above issues, we propose a delay-tolerant CSMA-based MAC scheme, named UAV delay-tolerant multiple access control (UD-MAC), that provides more access opportunities for SCF mode by setting the priority of SCF mode. It satisfies the dynamic performance requirement of nodes that in SCF and MH mode modes by an adaptive freezing period. For clarity, we summarize the main contributions as follows.
\begin{itemize}
	\item[1.] We consider both the MH mode and the SCF mode in MAC. The SCF mode provides nodes with high throughput links once they meet neighbors that are returning to ground unit (GU), and the MH mode allows nodes that do not have SCF opportunities to forward data to the GU.
	\item[2.] We design a novel MAC protocol, that could 1) assign more resource occupation rights to nodes in SCF mode by the priority of access, and 2) freeze the time to offer more resources to the nodes in MH mode. To the best of our knowledge, this is the first MAC protocol that can dynamically satisfy the requirement of nodes in SCF and MH modes.
	\item[3.] We derive the closed-form solution of the probability of being in SCF mode, and further offer a closed-form solution of the network throughput. It reveals the connection between mobility and throughput and accurately describes the performance level of the given network. To meet the dynamic performance requirements of nodes, it can be used as a reference value for the throughput of SCF and MH modes to provide a benchmark for adaptive freezing period adjustment.
\end{itemize}

It is noted that parts of this paper have been published in our conference paper \cite{UD-MAC}. Compared with the conference version, this paper further generalizes SCF to CSMA and studies SCF and MH modes to improve the throughput of UWSN. Furthermore, the derivation of the closed-form solution of the throughput is based on the probability of being in the SCF mode in three dimensions, which is more realistic than the conference version. We propose to use SCF priority and a freezing period to dynamically meet the requirement of nodes in SCF and MH modes.

The remaining parts of this paper are organized as follows. Related works are discussed in Section II. Section III presents the UAV sensing network system model and the specific design of UD-MAC. The probability of being in SCF mode is from Section IV. In Section V, the network throughput is calculated. Section VI presents the numerical results to verify the theoretical results. Finally, Section VII concludes this paper. The key parameters and abbreviations are listed in \textcolor[rgb]{0,0,0}{Table \ref{tab1}}.

\begin{table}[]
	\caption{\label{tab1}Key parameters and abbreviations}
	\begin{center}
		\begin{tabular}{m{2cm}<{\centering}|m{6cm}<{\centering}}
			\hline
			\hline
			{\textbf{Symbol}} & {\textbf{Description}} \\
			\hline
			SCF & Store-carry-and-forward\\
			\hline
			MH & Multi-hop\\
			\hline
			CSMA/CA & Carrier Sense Multiple Access with Collision Avoid\\
			\hline
			$t$ & Waiting time\\
			\hline
			$R$ & Radius of the scene\\
			\hline
			$r$ & Communication range\\
			\hline
			$H$ & height of UAVs' 1-D or 2-D space of activities from the ground\\
			\hline
			$\bar{v}$ & UAVs' speed\\
			\hline
			$d$ & Distance from the UAV to GU\\
			\hline
			$p_{t}\left(x, y, z\right)$ & the probability that a non-returning UAV located at $\left(x, y, z\right)$ is in SCF mode within the waiting time $t$\\
			\hline
			$N$ & Number of UAVs within two hops\\
			\hline
			$N_{s c f}$ & Number of UAVs being in SCF mode within two hops\\
			\hline
			$S$ & Network throughput\\
			\hline
			$M$ & Number of data channels\\
			\hline
			$E[P]$ & The average packet payload size\\
			\hline
			\hline
		\end{tabular}
	\end{center}
\end{table}

\section{related works}
\label{sec:related works}
In the study of MAC protocols for distributed networks, time slot-based \cite{rela_01} and carrier sense multiple access with collision avoidance (CSMA/CA)-based \cite{rela_02} MAC protocols emerged originally.
In time division multiple access based reliable broadcast mechanism, vehicular MAC protocol (VeMAC) \cite{rela_1} divides time slots according to the vehicle driving direction for collision prediction and avoidance to ensure reliable and fast transmission. In \cite{rela_2}, Li \textit{et al.} used the assistance of neighboring vehicles to ensure the successful competition in the event of an access collision. In multi-hop (MH) wireless networks, Zhang \textit{et al.} proposed to select nodes with idle or less buffers as relay nodes to improve the network throughput in \cite{rela_3}. However, the time slot based schemes requires the pre-assignment at base station, which incurs additional signaling overhead. In addition, when there are multiple competing nodes, they are likely to choose the same time slot to send data, which is prone to competitive collisions, resulting in a waste of resources. CSMA/CA is a promising solution in UWSN owing to its conflict avoidance and asynchronous access strategies \cite{rela_4}. In \cite{rela_5}, Baek \textit{et al.} achieved higher throughput and lower end-to-end delay than CSMA/CA by introducing time mirroring. Kwon \textit{et al.} combined CSMA/CA schemes in uplink non-orthgonal multiple access to reduce the probability of packet collisions in \cite{rela_6}. But research on CSMA/CA is almost focused on how to improve network throughput by reducing collisions and hardly take into account UAV's high mobility, which is an essential feature that differs from traditional mobile networks. Furthermore, a data packet is usually forwarded multiple times in the MH network. The more nodes access the network, the greater competitive pressure on network resources, which worsens the performance of CSMA/CA.

It should be highlighted that there are generally several returning UAVs that moving back to the ground unit (GU) for charging owing to the limited energy. This provides a precise opportunity to inspire the store-carry-and-forward (SCF) transmission mode.
In high mobility networks, SCF transmission mode is originally proposed to reduce transmission outage time \cite{scf_1}\cite{scf_2}. The SCF mode can forward data through the mobility of nodes, which solves the problem of intermittent connection and low throughput \cite{scf_3}. Many scholars have also found that the SCF mode is very helpful to reduce energy consumption.
Zhang \textit{et al.} proposed a SCF-based scheme to achieve energy gain by more than 70\% compared with the MH-only scheme \cite{scf_4}.
A similar conclusion, namely a 70\% reduction of energy consumption, is realized by dynamically shifting the network mode between SCF and MH \cite{scf_5}.
In addition to the superiority of energy consumption, the SCF mode also has the ability to increase the capacity by $\Theta((n /(\log n)))$ times in the UWSN with $n$ UAV nodes \cite{scf_6}.
In \cite{scf_7}, Zheng \textit{et al.} proposed a SCF scheme based on random sub-channel selection to improve system throughput and energy efficiency.
Another SCF-based MAC protocol, which divided the time slot into two parts for the access of returning UAVs and non-returning UAVs, is proposed in \cite{UD-MAC}.
However, the propensity of resources to UAVs in SCF mode is ignored, which makes it difficult to guarantee the performance of SCF mode. Besides, the above research about SCF also omits the impact of nodes’ mobility on network performance.

However, we try to find ways to improve network throughput from the characteristics of UAV's mobility and limited energy and propose to support SCF and MH modes in MAC. In addition, we also consider the resource skewing problem in SCF mode and propose the first MAC protocol that can dynamically meet the needs of nodes in SCF and MH mode.

\section{Multiple Access Protocol design}
\label{sec:protocol}
In this section, we first illustrate the transmission process in the SCF and MH modes with examples in the system model. Then, the time and frequency resources required for the transmission process are divided. Finally, the access mechanism of UD-MAC is described, focusing on the access priority of SCF and MH modes and the freezing period used to solve the resource skewing problem.
\subsection{System Model}
\begin{figure}[t]
	\centering
	\includegraphics[width=0.3\textheight]{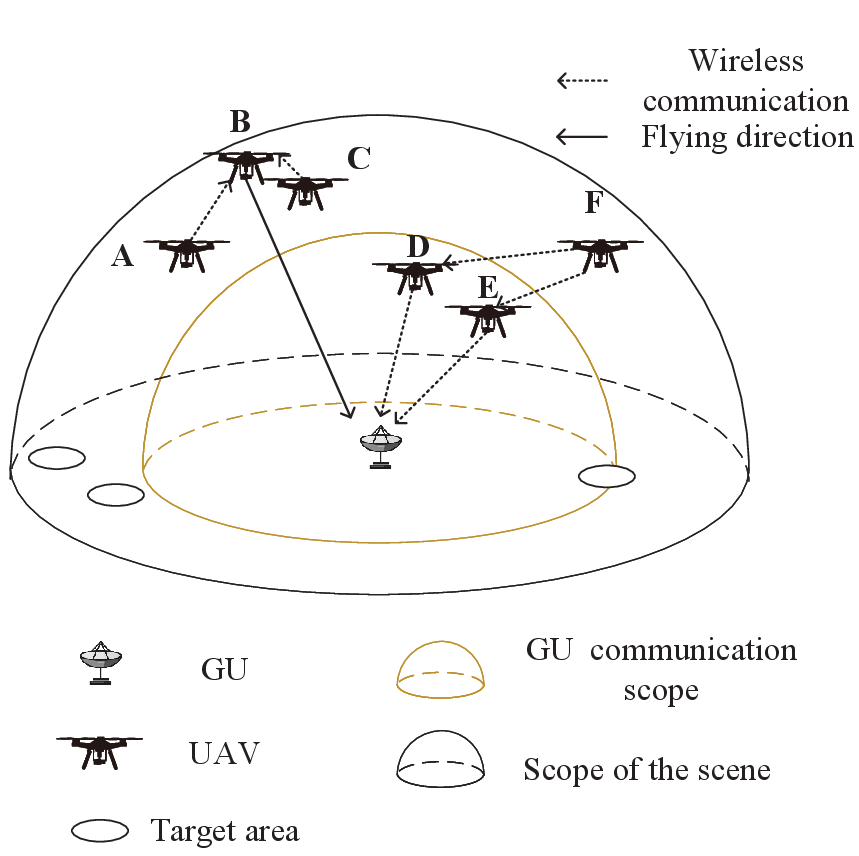}
	\DeclareGraphicsExtensions.
	\caption{Coexistence of SCF and MH transmission modes in UWSN.}
	\label{fig 1}
\end{figure}

The system model of the UWSN consists mainly of several UAVs and a GU. The GU is responsible for controlling UAVs and collecting data from UAVs. UAVs are mainly located in a three-dimensional (3-D) space centered on the GU. UAVs have a limited battery capacity, resulting in short flight times. Therefore, they often fly back to GU for recharging. When the UAV returns to GU, it can store and carry other UAVs' data along the return path. In this way, the wasted returning time can be used to increase the capacity of the UAV sensing network. As shown in \textcolor[rgb]{0,0,0}{Fig. \ref{fig 1}}, when UAV A has data to send to GU, it forwards its data to UAV B, which returns through UAV A. Then, the data is transmitted from UAV B to GU. However, UAV B does not necessarily appear within the communication range of UAV A at first. Hence, it is possible that UAV A hovered for some time to wait for UAV B to pass by. In addition, UAV F can communicate with GU via UAV D or UAV E in MH transmission mode.

\subsection{Channel Multiplexing}
\begin{figure}[t]
	\centering
	\includegraphics[width=0.32\textheight]{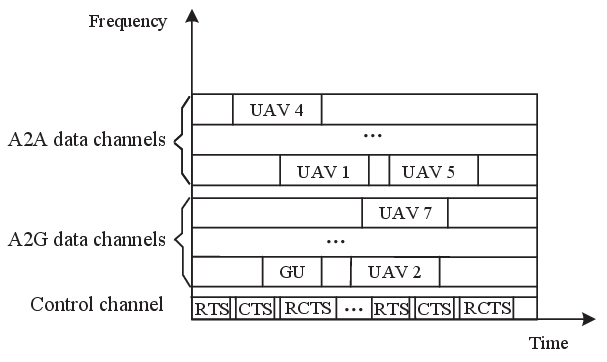}
	\DeclareGraphicsExtensions.
	\caption{Time and frequency resource division.}
	\label{fig 2}
\end{figure}

As shown in \textcolor[rgb]{0,0,0}{Fig. \ref{fig 2}}, the entire spectrum resource is divided into data channels for transmitting data packets and a control channel for transmitting control packets. The data channels are divided into air-to-air (A2A) data channels and air-to-ground (A2G) data channels. Therefore, the data transmission between UAVs does not affect the data transmission from UAVs to GU. All UAVs and GU share one control channel, and compete data channels' time and frequency resources by sending control packets including require for service (RTS), clear to send (CTS), and repeat CTS (RCTS).

\subsection{UD-MAC Access Mechanism}
\label{UD-MAC}
\begin{figure}[t]
	\centering
	\includegraphics[width=0.35\textheight]{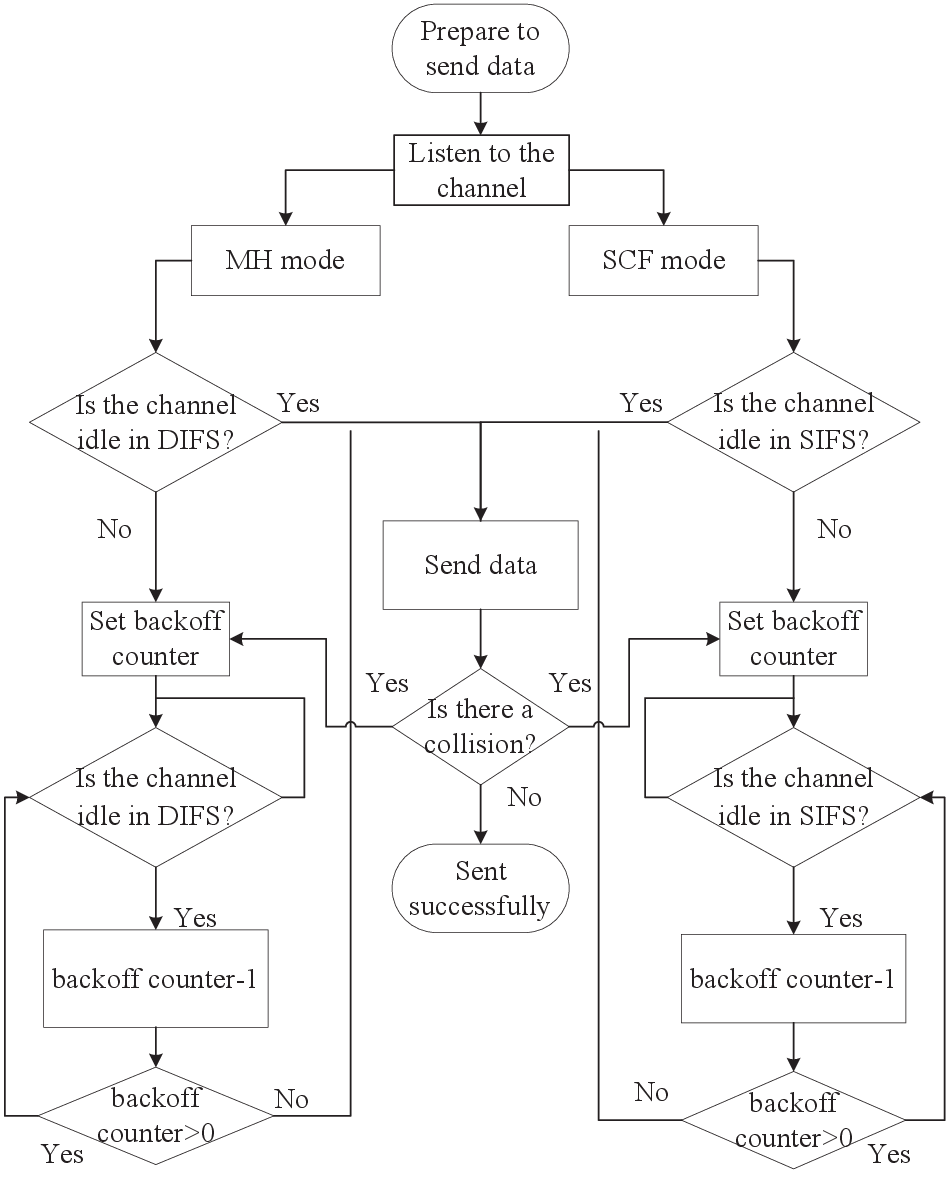}
	\DeclareGraphicsExtensions.
	\caption{Control channel access mechanism.}
	\label{fig 3}
\end{figure}

As shown in \textcolor[rgb]{0,0,0}{Fig. \ref{fig 3}}, when the UAV wants to send data, it will first listen to the control channel. If the control channel is idle, it will send an RTS. Otherwise, it will set the back-off counter randomly. The back-off counter will only count down when the control channel is idle. When the backoff counter is 0, it can try to send an RTS again.

SCF mode utilizes the mobility of UAVs to reduce access times, which can reduce competition in the network. Therefore, our scheme sets UAVs in SCF mode to have a higher priority to access the control channel than UAVs in MH mode. If the UAV is in MH mode, it needs to wait for a distributed inter-frame spacing (DIFS) before sending an RTS. If the control channel is not occupied during DIFS, the UAV can send an RTS. If the UAV is in SCF mode, there is only a short inter-frame space (SIFS) to wait for before sending an RTS. Since DIFS is longer than SIFS, the access to the control channel for the UAV waiting for the DIFS interval will be seized by the UAV waiting for the SIFS interval. Hence, UAVs in SCF mode only need to avoid conflicts with UAVs that are also in SCF mode, while UAVs in MH mode need to avoid conflicts with all UAVs.

\begin{figure}[t]
	\centering
	\includegraphics[width=0.23\textheight]{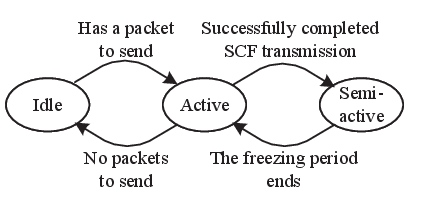}
	\DeclareGraphicsExtensions.
	\caption{The three states of UAVs.}
	\label{fig 4}
\end{figure}

Since UAVs in SCF mode have a higher access priority, it may cause a skewing of network resources. We balance the trade-off the resources of MH mode and SCF mode by setting a freezing period. When the UAV successfully completes SCF mode transmission, it needs to enter a freezing period,  which references the benchmark provided by the throughput of the SCF and MH modes. During this period, the UAV acts as a receiver only and does not actively compete for channel resources. The three states of UAVs shown in \textcolor[rgb]{0,0,0}{Fig. \ref{fig 4}} are idle, active, and semi-active. In the idle state, the UAV has no packets to send. In the active state, the UAV tries to access the control channel. The semi-active state means that the UAV does not actively compete for access to the control channel, but can receive data transmitted by other UAVs. When the freezing period ends, the UAV becomes active again.

\section{The Probability of Being in SCF Mode}
\label{sec:probility}
In this section, we mainly calculate the probability of being in SCF mode for one-dimensional (1-D), two-dimensional (2-D) and 3-D, and consider the effect of node position and waiting time on it. The probability of being in SCF mode is the probability that a non-returning UAV can encounter a returning UAV, by which the ratio of the number of UAVs in SCF mode to the total number of UAVs in a small area can be calculated. Therefore, the probability of being in SCF mode is required to analyze the throughput of a network where SCF mode and MH mode coexist. Instead of directly setting the probability of being in SCF mode to a certain value, we use geometric graphs to derive the probability of being in SCF mode, aiming to provide valid theoretical support for the subsequent throughput analysis. In particular, we consider the waiting time to reflect the dynamicity of the network. The probability of being in SCF mode is affected by the waiting time, during which a non-returning UAV can hover to wait for a returning UAV to appear in its communication range.

\begin{figure}[t]
	\centering
	\includegraphics[width=0.34\textheight]{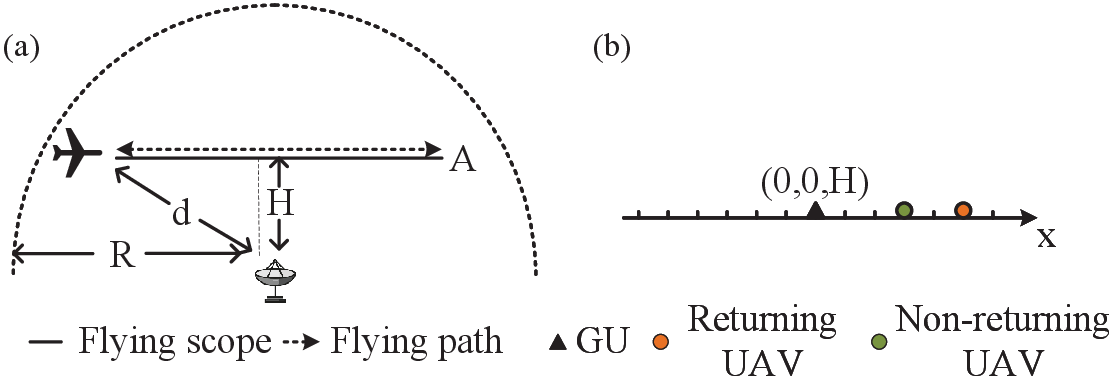}
	\DeclareGraphicsExtensions.
	\caption{UAVs' 1-D space of activities. (a) the scene diagram of the UAV's activity range in 1-D. (b) the top-view geometric diagram of the UAV's activity range in 1-D}
	\label{fig 5}
\end{figure}

\begin{figure}[t]
	\centering
	\includegraphics[width=0.34\textheight]{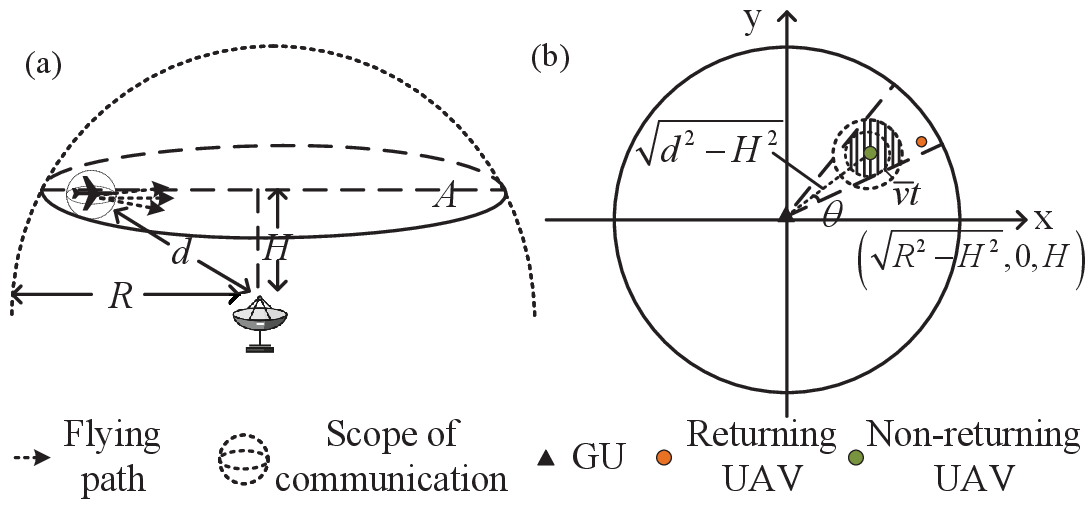}
	\DeclareGraphicsExtensions.
	\caption{UAVs' 2-D space of activities. (a) the scene diagram of the UAV's activity range in 2-D. (b) the top-view geometric diagram of the UAV's activity range in 2-D.}
	\label{fig 6}
\end{figure}

When UAVs perform different tasks, such as pipeline inspection, marine surveillance, and forest inspection, their space of activities can be divided into one-dimensional (1-D) space, two-dimensional (2-D) space, and 3-D space. \textcolor[rgb]{0,0,0}{Fig. \ref{fig 5}(a)}, \textcolor[rgb]{0,0,0}{Fig. \ref{fig 6}(a)} and \textcolor[rgb]{0,0,0}{Fig. \ref{fig 7}(a)} depict the scene diagrams of the UAV's activity range in three dimensions. In order to calculate the probability of being SCF mode, we abstract the scene graphs into geometric graphs, as shown in \textcolor[rgb]{0,0,0}{Fig. \ref{fig 5}(b)}, \textcolor[rgb]{0,0,0}{Fig. \ref{fig 6}(b)} and \textcolor[rgb]{0,0,0}{Fig. \ref{fig 7}(b)}. In particular, \textcolor[rgb]{0,0,0}{Fig. \ref{fig 6}(b)} is a top-view geometric diagram of \textcolor[rgb]{0,0,0}{Fig. \ref{fig 6}(a)}.
Define the radius of the scene as $R$, and the communication range of inter-UAV and UAV-GU as $r$. The height of UAVs' 1-D or 2-D space of activities from the ground is $H$. The UAVs fly at a constant speed of $\bar{v}$. $d$ is the distance from the UAV to GU. In the subsequent analysis in this section, the UAVs that request a returning UAV to carry packets are referred to as non-returning UAVs and are distributed in space outside the GU's communication range and within the range of the hemispheric scenario. To achieve high-capacity transmission, we assume that the returning UAV is still flying in the active range of UAVs when it is not in the communication range of GU to provide SCF transmission for the non-returning UAVs, and only flies back to GU with the shortest distance after entering the communication range of GU.

When the UAVs are sent sequentially in the time dimension, they will return at any time as energy is consumed. As shown in \textcolor[rgb]{0,0,0}{Fig. \ref{fig 8}}, the returning UAV in the red region can forward the data from the non-returning UAV to GU, so that the probability of the non-returning UAV being in SCF mode is related to the ratio of the red region to the space of activities. It is worth noting that the red region is related to the waiting time $t$. As $t$ increases, the red region cannot increase indefinitely because the returning UAVs cannot appear outside the scene boundary. Therefore, two cases are considered to calculate the probability of being in SCF mode. Case I is that the red region does not exceed the scene boundary, and case II is that the red region exceeds the scene boundary.


\begin{figure*}[htpb]
	\begin{equation}
	\label{equ 1}
p_{t}\left(\sqrt{d^{2}-H^{2}}, 0, H\right)= \begin{cases}\frac{2 r+\bar{v} t}{2\left(\sqrt{R^{2}-H^{2}}-\sqrt{r^{2}-H^{2}}\right)}, & r+\bar{v} t \leq \sqrt{R^{2}-H^{2}}-\sqrt{d^{2}-H^{2}} \\ \frac{r+\sqrt{R^{2}-H^{2}}-\sqrt{d^{2}-H^{2}}}{2\left(\sqrt{R^{2}-H^{2}}-\sqrt{r^{2}-H^{2}}\right)}, & \sqrt{R^{2}-H^{2}}-\sqrt{d^{2}-H^{2}}<r+\bar{v} t\end{cases}.
\end{equation}
\end{figure*}

\newcounter{Temp1} 
\setcounter{equation}{7} 
\begin{figure*}[ht]
\begin{equation}
	\label{equ 2}
p_{t}(x, y, \mathrm{H})= \begin{cases}\frac{\left(\frac{1}{2}-\frac{\theta}{\pi}\right) \pi r^{2}+r(r+\bar{v} t) \sin \alpha}{\pi\left(R^{2}-r^{2}\right)}+\frac{\left(\frac{\pi}{2}+\theta-\alpha\right)(r+\bar{v} t)^{2}}{\pi\left(R^{2}-r^{2}\right)}, & r+\bar{v} t \leq \sqrt{R^{2}-H^{2}}-\sqrt{d^{2}-H^{2}} \\ \frac{\left(\frac{1}{2}-\frac{\theta}{\pi}\right) \pi r^{2}+\theta\left(R^{2}-H^{2}\right)}{\pi\left(R^{2}-r^{2}\right)}-\frac{r \sqrt{d^{2}-H^{2}} \cos \theta}{\pi\left(R^{2}-r^{2}\right)}, & \sqrt{R^{2}-H^{2}}-\sqrt{d^{2}-H^{2}}<r+\bar{v} t\end{cases}.
\end{equation}
\end{figure*}
\setcounter{equation}{\value{Temp1}}

\begin{figure}[t]
	\centering
	\includegraphics[width=0.34\textheight]{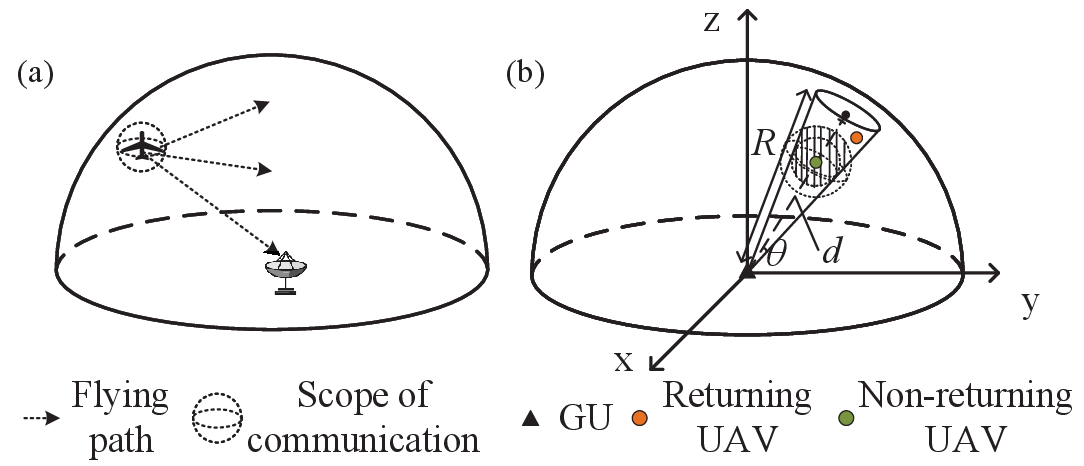}
	\DeclareGraphicsExtensions.
	\caption{UAVs' 3-D space of activities. (a) the scene diagram of the UAV's activity range in 3-D. (b) the geometric diagram of the UAV's activity range in 3-D}
	\label{fig 7}
\end{figure}

\subsection{One-dimensional}
\begin{figure}[!htbp]
	\centering
	\subfigure[{Case I.}]{
		\label{fig 8(a)}
		\includegraphics[width=0.21\textwidth]{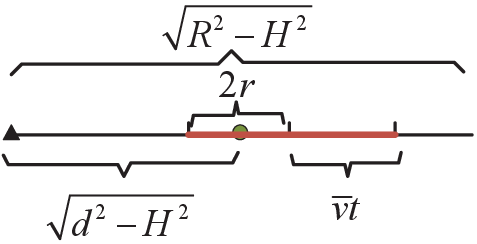}}
	\subfigure[{Case II.}]{
		\label{fig 8(b)}
		\includegraphics[width=0.25\textwidth]{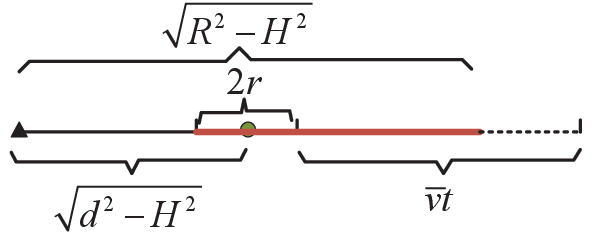}}
	\caption{returning UAVs in 1-D activities space can appear in the neighborhood of the non-returning UAV within $t$. }
	\label{fig 8}
\end{figure}

Because UAVs are active on the x-axis, the position of the non-returning UAV can be represented by $\left(\sqrt{d^{2}-H^{2}}, 0, H\right)$. As shown in \textcolor[rgb]{0,0,0}{Fig. \ref{fig 8(a)}}, when $t$ satisfies $r+\bar{v} t \leq \sqrt{R^{2}-H^{2}}-\sqrt{d^{2}-H^{2}}$, the length of the red region is $r+\bar{v} t$. As shown in \textcolor[rgb]{0,0,0}{Fig. \ref{fig 8(b)}}, when $\sqrt{R^{2}-H^{2}}-\sqrt{d^{2}-H^{2}}<r+\bar{v} t$, the length of the red region is determined by the distance from the non-returning UAV to the scene boundary, which is $r+\sqrt{R^{2}-H^{2}}-\sqrt{d^{2}-H^{2}}$.

Therefore, the probability that a non-returning UAV located at $\left(\sqrt{d^{2}-H^{2}}, 0, H\right)$ is in SCF mode within the waiting time $t$ is given by \eqref{equ 1}.

\subsection{Two-dimensional}
\begin{figure}[!htbp]
	\centering
	\subfigure[{Case I.}]{
		\label{fig 9(a)}
		\includegraphics[width=0.21\textwidth]{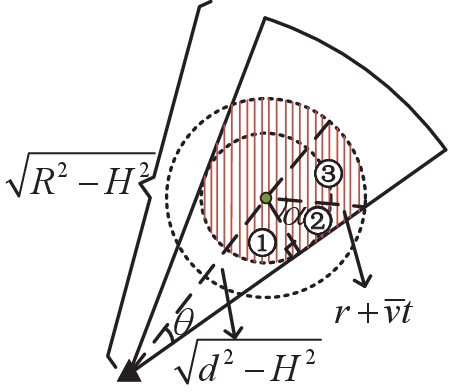}}
	\subfigure[{Case II.}]{
		\label{fig 9(b)}
		\includegraphics[width=0.21\textwidth]{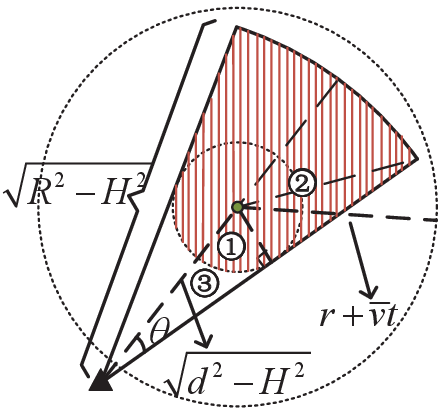}}
	\caption{returning UAVs in 2-D activities space can appear in the neighborhood of the non-returning UAV within $t$. }
	\label{fig 9}
\end{figure}

In a 2-D space of activities, we calculate the red region for both cases to obtain the probability of being in SCF mode, as shown in \textcolor[rgb]{0,0,0}{Fig. \ref{fig 9(a)}} and \textcolor[rgb]{0,0,0}{Fig. \ref{fig 9(b)}}. The position of the non-returning UAV is set to $(x, y, \mathrm{H})$.

\begin{itemize}
\item[$\bullet$]When $r+\bar{v} t \leq \sqrt{R^{2}-H^{2}}-\sqrt{d^{2}-H^{2}}$, we need to calculate the red region in \textcolor[rgb]{0,0,0}{Fig. \ref{fig 9(a)}}. Due to the symmetrical red region, we only need to calculate the sum of regions \normalsize{\textcircled{\scriptsize{1}}}\normalsize, \normalsize{\textcircled{\scriptsize{2}}}\normalsize, and \normalsize{\textcircled{\scriptsize{3}}}\normalsize.

The area of region \normalsize{\textcircled{\scriptsize{1}}}\normalsize\enspace is 
\setcounter{equation}{1}
\begin{equation}
	\label{2d1}
S_{1}=\frac{\pi / 2-\theta}{2 \pi} \pi r^{2}=\left(\frac{1}{4}-\frac{\theta}{2 \pi}\right) \pi r^{2},
\end{equation}
where $\sin \theta=\frac{r}{\sqrt{d^{2}-H^{2}}}$.

The area of region \normalsize{\textcircled{\scriptsize{2}}}\normalsize\enspace is
\begin{equation}
	\label{2d2}
S_{2}=\frac{1}{2} r(r+\bar{v}) \sin a,
	\end{equation}
where $\cos a=\frac{r}{r+\bar{v} t}$.

The area of region \normalsize{\textcircled{\scriptsize{3}}}\normalsize\enspace is
\begin{equation}
	\label{2d3}
S_{3}=\frac{\pi / 2+\theta-a}{2 \pi} \pi(r+\bar{v} t)^{2}.
\end{equation}
	
Thus, the area of the red region is $2(S_{1} + S_{2} + S_{3})$ by \eqref{2d1}, \eqref{2d2}, and \eqref{2d3}.

\item[$\bullet$]When $\sqrt{R^{2}-H^{2}}-\sqrt{d^{2}-H^{2}}<r+\overline{v t}$, we divide the sector region into three parts: regions \normalsize{\textcircled{\scriptsize{1}}}\normalsize, \normalsize{\textcircled{\scriptsize{2}}}\normalsize, and \normalsize{\textcircled{\scriptsize{3}}}\normalsize. But ultimately we only need regions \normalsize{\textcircled{\scriptsize{1}}}\normalsize\enspace and \normalsize{\textcircled{\scriptsize{2}}}\normalsize\enspace to get the red region.

The area of region \normalsize{\textcircled{\scriptsize{2}}}\normalsize\enspace is
\begin{equation}
	\label{2d4}
S_{1}=\frac{\pi / 2-\theta}{2 \pi} \pi r^{2}=\left(\frac{1}{4}-\frac{\theta}{2 \pi}\right) \pi r^{2},
\end{equation}
where $\sin \theta=\frac{r}{\sqrt{d^{2}-H^{2}}}$.

The sum of regions \normalsize{\textcircled{\scriptsize{1}}}\normalsize, \normalsize{\textcircled{\scriptsize{2}}}\normalsize, and \normalsize{\textcircled{\scriptsize{3}}}\normalsize\enspace is
\begin{equation}
	\label{2d5}
S_{2}=\frac{\theta}{2 \pi} \pi\left(R^{2}-H^{2}\right).
\end{equation}

The sum of regions \normalsize{\textcircled{\scriptsize{1}}}\normalsize\enspace and \normalsize{\textcircled{\scriptsize{3}}}\normalsize\enspace is
\begin{equation}
	\label{2d6}
S_{3}=\frac{1}{2} r \sqrt{d^{2}-H^{2}} \sin \left(\frac{\pi}{2}-\theta\right)=\frac{1}{2} r \sqrt{d^{2}-H^{2}} \cos \theta.
\end{equation}
	
Thus, the area of red region is $2(S_{1} + S_{2} - S_{3})$ by \eqref{2d4}, \eqref{2d5}, and \eqref{2d6}.
\end{itemize}

Therefore, the probability that a non-returning UAV located at $(x, y, \mathrm{H})$ is in SCF mode within the waiting time $t$ is given by \eqref{equ 2}.


\setcounter{equation}{14}
\begin{figure*}[htpb]
\begin{equation}
	\label{equ 3}
p_t(x, y, z)=\left\{\begin{array}{l}
\frac{r^3\left(1-\frac{r}{d}\right)^2\left(2+\frac{r}{d}\right)}{2\left(R^3-r^3\right)}+\frac{(r+\bar{v} t)^3[1-\sin (\alpha-\theta)]^2[2+\sin (\alpha-\theta)]}{2\left(R^3-r^3\right)}+ \\
\frac{[r \sin \theta+(r+\bar{v} t) \sin (\alpha-\theta)]}{2\left(R^3-r^3\right)} \times \frac{\left[(r \cos \theta)^2+(r+\bar{v} t)^2 \cos^2 (\alpha-\theta)+r(r+\bar{v} t) \cos \theta \cos (\alpha-\theta)\right]}{2\left(R^3-r^3\right)}, r+\bar{v} t \leq R-d \\
\frac{2 R^3 d^2\left(d-\sqrt{d^2-r^2}\right)+2 r^3 d^3-r^4 d^2 - r^2 d^4}{2d^3\left(R^3-r^3\right)}, R-d<r+\bar{v} t
\end{array}\right.
.
\end{equation}
\end{figure*}

\subsection{Three-dimensional}
\begin{figure}[!htbp]
	\centering
	\subfigure[{Case I.}]{
		\label{fig 10(a)}
		\includegraphics[width=0.18\textwidth]{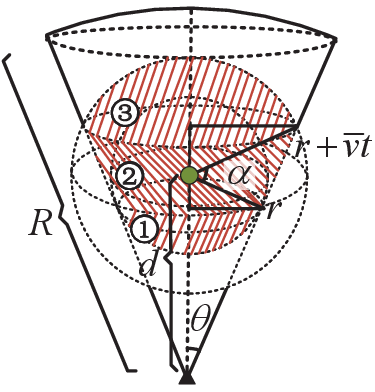}}
	\subfigure[{Case II.}]{
		\label{fig 10(b)}
		\includegraphics[width=0.22\textwidth]{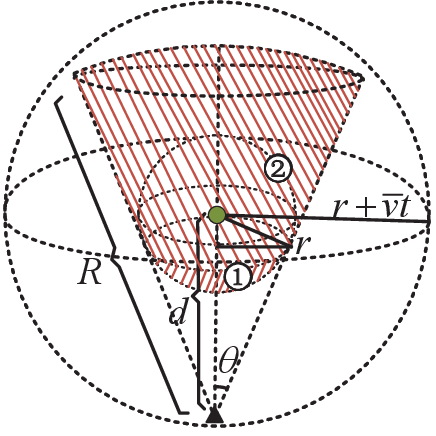}}
	\caption{returning UAVs in 3-D activities space can appear in the neighborhood of the non-returning UAV within $t$. }
	\label{fig 10}
\end{figure}

In the 3-D space of activities, we will calculate the volume of the red region. The coordinates of the non-returning UAV is defined as $(x,y,z)$, as shown in \textcolor[rgb]{0,0,0}{Fig. \ref{fig 10}}.

\begin{itemize}
\item[$\bullet$]When $r+\bar{v} t \leq R-d$, the volume of the red region including regions \normalsize{\textcircled{\scriptsize{1}}}\normalsize, \normalsize{\textcircled{\scriptsize{2}}}\normalsize, and \normalsize{\textcircled{\scriptsize{3}}}\normalsize\enspace in \textcolor[rgb]{0,0,0}{Fig. \ref{fig 10(a)}} is calculated.

The volume of region \normalsize{\textcircled{\scriptsize{1}}}\normalsize\enspace is
\setcounter{equation}{8}
\begin{equation}
	\label{3d1}
V_{1}=\frac{1}{3} \pi r^{3}\left(1-\frac{r}{d}\right)^{2}\left(2+\frac{r}{d}\right),
	\end{equation}
where $\sin \theta=\frac{r}{d}$.

The volume of region \normalsize{\textcircled{\scriptsize{2}}}\normalsize\enspace is
\begin{equation}
	\label{3d2}
\begin{gathered}
V_2=\frac{1}{3} \pi[r \sin \theta+(r+\bar{v} t) \sin (\alpha-\theta)] \times \\
\quad\quad\quad{\left[(r \cos \theta)^2+(r+\bar{v} t)^2 \cos ^2(\alpha-\theta)+\right.} \\
r(r+\bar{v} t) \cos \theta \cos (\alpha-\theta)]
\end{gathered},
\end{equation}
where $\cos \alpha=\frac{r}{r+\bar{v} t}$.

The volume of region \normalsize{\textcircled{\scriptsize{3}}}\normalsize\enspace is
\begin{equation}
	\label{3d3}
V_{3}=\frac{1}{3} \pi(r+\bar{v} t)^{3}[1-\sin (\alpha-\theta)]^{2}[2+\sin (\alpha-\theta)].
\end{equation}

Thus, the volume of the red region is $(V_{1} + V_{2} + V_{3})$ by \eqref{3d1}, \eqref{3d2}, and \eqref{3d3}.
\end{itemize}

\begin{itemize}
\item[$\bullet$]When $R-d<r+\overline{v t}$, the volume of the red region including regions \normalsize{\textcircled{\scriptsize{1}}}\normalsize\enspace and \normalsize{\textcircled{\scriptsize{2}}}\normalsize\enspace in \textcolor[rgb]{0,0,0}{Fig. \ref{fig 10(b)}} is calculated.
The volume of region \normalsize{\textcircled{\scriptsize{1}}}\normalsize\enspace is
\begin{equation}
	\label{3d4}
V_{1}=\frac{1}{3} \pi r^{3}\left(1-\frac{r}{d}\right)^{2}\left(2+\frac{r}{d}\right).
\end{equation}

The volume of region \normalsize{\textcircled{\scriptsize{2}}}\normalsize\enspace is
\begin{equation}
	\label{3d5}
	V_2=\frac{1}{3} \pi \frac{2 R^3 d^2\left(d-\sqrt{d^2-r^2}\right)-r^2\left(d^2-r^2\right)^2}{d^3}
\end{equation}
Thus, the volume of the red region is $(V_{1} + V_{2})$ by \eqref{3d4} and \eqref{3d5}.
\end{itemize}

Therefore, the probability that a non-returning UAV located at $(x, y, z)$ is in SCF mode within the waiting time $t$ is given by \eqref{equ 3}.

Above, we calculated the probability of being in SCF mode for 1-D, 2-D, and 3-D. Based on the probability that the non-returning UAV is in SCF, we can analyze the proportion of data transmitted using SCF mode and using MH mode in the whole network, and thus analyze the network throughput.

\section{Analysis of Network Throughput}
\label{sec:throughput}
\begin{figure}[!htpb]
	\centering
	\includegraphics[width=0.35\textheight]{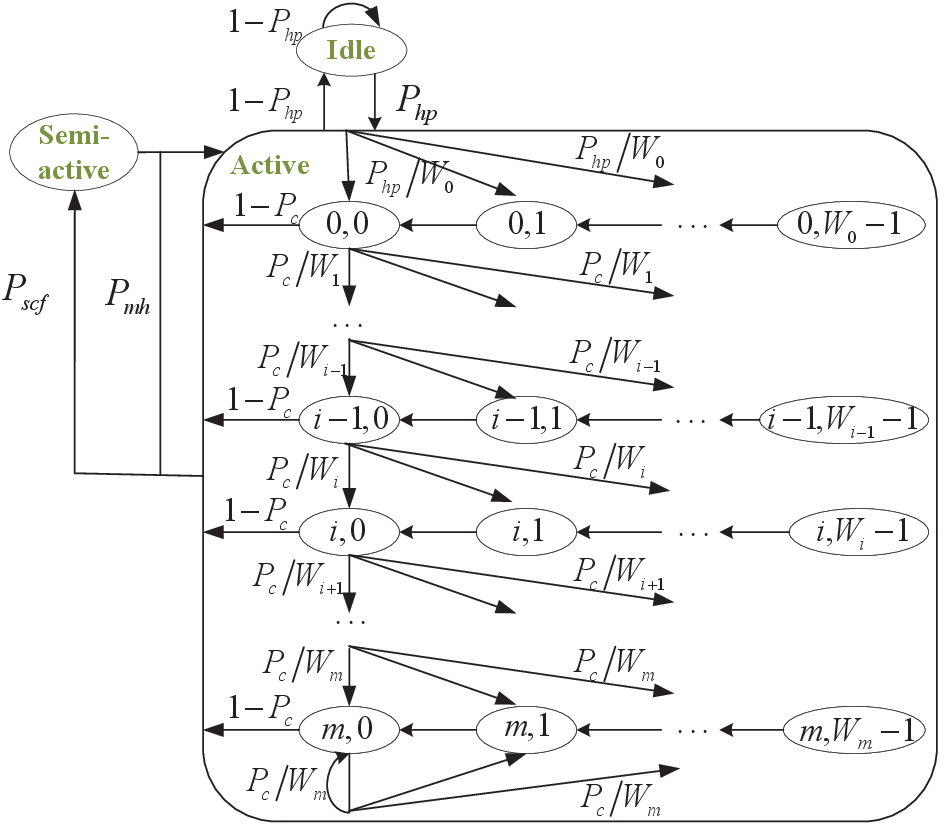}
	\DeclareGraphicsExtensions.
	\caption{The discrete Markov state model.}
	\label{fig 11}
\end{figure}

In this section, we analyze the network saturation throughput based mainly on the probability of being in SCF mode in Section IV to measure the performance of UD-MAC. We can get the number of UAVs in SCF mode and the number of UAVs in MH mode by the probability of being in SCF mode, and then analyze the conflict probability and successful transmission probability. First, we build a discrete Markov state model to describe the states of the UAV in SCF mode and the UAV in MH mode. As shown in \textcolor[rgb]{0,0,0}{Fig. \ref{fig 11}}, the three circles in the upper left corner correspond to the three states mentioned in Section \ref{UD-MAC}, namely idle, active, and semi-active states. The other circles depict the states in which the active UAV has a backoff counter with value $b(t_{0})$ at the $s(t_{0})$th attempt to access the control channel at moment $t_{0}$, denoted by $(s(t_{0}),b(t_{0}))$. In the steady-state condition, $b_{i, j}=\lim _{t_{0} \rightarrow \infty} p\left\{s(t_{0})=i, b(t_{0})=j\right\}$. We can derive the network throughput $S$ from $b_{i, j}$.

For convenience, define $W=CW_{min}$ as the minimum backoff window. Let $m$ be the maximum backoff stage such that $W_{max}=2^mW$ is the maximum backoff window. Let us define the backoff window at the $i$th backoff as the symbol of $W_i=2^iW$, where $i \in[0, m]$. The value of the UAV's backoff counter for each backoff is a randomly selected number from $[0, W_i - 1]$ for the countdown. We set the probability that the UAV has a packet to send to $P_{hp}$. The $P_c$ will be referred to as the independent collision probability. The probability of being in SCF transmission mode is $P_{scf}$ and the probability of being in MH transmission mode is $P_{mh}$. As shown in \textcolor[rgb]{0,0,0}{Fig. \ref{fig 11}}, we describe the transfer probabilities between each state using the four probabilities $P_{hp}$, $P_c$, $P_{scf}$, and $P_{mh}$.

Next, we easily obtain the Markov state one-step transfer probability in a steady state with
\setcounter{equation}{14}
\begin{equation}
	\label{eq 15}
\left\{\begin{array}{l}
b_{i, 0}=P_c^i b_{0,0}, 0<i<m \\
b_{m, 0}=\frac{P_c^m}{1-P_c} b_{0,0} \\
b_{s-a c t i v e}=P_{s c f} b_{0,0} \\
b_{\text {idle }}=\frac{1-P_{h p}}{P_{h p}} \\
b_{i, k}=\frac{W_i-k}{W_i} \cdot\left\{\begin{array}{l}
(1-p) \sum_{j=0}^m b_{j, 0}, i=0 \\
p \cdot b_{i-1,0}, 0<i<m \\
p \cdot\left(b_{m-1,0}+b_{m, 0}\right), i=m
\end{array}\right.
\end{array}\right.
.
\end{equation}

According to \cite{throughout_1}, we can get

\begin{equation}
\label{eq 16}
\begin{aligned}
\sum_{i=0}^m \sum_{k=0}^{W_i-1} b_{i, k}= \frac{b_{0,0}\left(1-2 P_c\right)(W+1)+P_c W\left(1-\left(2 P_c\right)^m\right)}{2\left(1-2 P_c\right)\left(1-P_c\right)}
\end{aligned}.
\end{equation}

Thus, through \eqref{eq 15} and \eqref{eq 16}, all values of $b_{i,k}$ can be expressed as a function of $b_{0,0}$, which is ultimately determined by imposing the probability normalization condition, simplified as follows as 

\begin{equation}
\label{eq 17}
\begin{aligned}
1= & b_{0,0} \frac{\left(1-2 P_c\right)(W+1)+P_c W\left(1-\left(2 P_c\right)^m\right)}{2\left(1-2 P_c\right)\left(1-P_c\right)}+ \\
& P_{s c f} b_{0,0}+\frac{1-P_{h p}}{P_{h p}}
\end{aligned}.
\end{equation}

The probability that the UAV has a backoff counter of 0 and is ready to send data is $\tau$, represented by 
\begin{equation}
\label{eq 18}
\tau=\sum_{i=0}^{m} b_{i, 0}=\frac{b_{0,0}}{1-P_{c}}.
\end{equation}
Substituting \eqref{eq 17} into \eqref{eq 18} yields the expansion of $\tau$ as

\begin{equation}
\label{eq 19}
\tau=\frac{2\left(1-2 P_c\right)\left(2 P_{h p}-1\right)}{\left(\begin{array}{l}
P_{h p}\left(1-2 P_c\right)(W+1)+P_c W\left(1-\left(2 P_c\right)^m\right)+ \\
2 P_{h p}\left(1-2 P_c\right)\left(2 P_{h p}-1\right) P_{s c f}
\end{array}\right)}.
\end{equation}

According to \eqref{eq 19}, we find that we need to discuss the four probabilities $P_{hp}$, $P_c$, $P_{scf}$ and $P_{mh}$ to solve for $\tau$. To solve for the network saturation throughput, we assume that the UAV always has packets to send, so that $P_{hp}=1$. Because the UAV transmits data either in SCF mode or in MH mode, $P_{mh}=1-P_{scf}$. Assuming that the number of UAVs within two hops is $N$, the number of UAVs in the SCF transmission mode within two hops is $N_{scf}=\sum_{i=1}^{N} p_{t}\left(x_{i}, y_{i}, z_{i}\right)$, calculated in Section \ref{sec:probility}. 

When $P_{scf} = 1$, \textcolor[rgb]{0,0,0}{Fig. \ref{fig 11}} depicts the Markov state model of the UAV in SCF mode. $P_{c 1}$ is the collision probability of the UAV in SCF mode, and $\tau_{s c f}$ is the transmission probability of the UAV in SCF mode. A collision occurs when a UAV in SCF mode sends a packet if at least one UAV in SCF mode is also sending a packet. According to \eqref{eq 19}, it can be obtained

\begin{equation}
\label{eq 20_1}
\left\{\begin{array}{l}
\tau_{s c f}=\frac{2\left(1-2 P_{c 1}\right)\left(2 P_{h p}-1\right)}{\left(\begin{array}{l}
P_{h p}\left(1-2 P_{c 1}\right)(W+1)+ \\
P_c W\left(1-\left(2 P_{c 1}\right)^m\right)+ \\
2 P_{h p}\left(1-2 P_{c 1}\right)\left(2 P_{h p}-1\right)
\end{array}\right)} \\
P_{c 1}=1-\left(1-\tau_{s c f}\right)^{N_{s c f}-1}
\end{array}\right.
.
\end{equation}

When $P_{scf} = 0$, \textcolor[rgb]{0,0,0}{Fig. \ref{fig 11}} depicts the Markov state model of the UAV in MH mode. $P_{c 2}$ is the collision probability of the UAV in MH mode, and $\tau_{m h}$ is the transmission probability of the UAV in MH mode. A collision occurs when a UAV in MH mode sends a packet if at least one UAV in MH mode is also sending a packet. According to \eqref{eq 19}, it can be obtained

\begin{equation}
\label{eq 20_2}
\left\{\begin{array}{l}
\tau_{m h}=\frac{2\left(1-2 P_{c 2}\right)\left(2 P_{h p}-1\right)}{P_{h p}\left(1-2 P_{c 2}\right)(W+1)+P_{c 2} W\left(1-\left(2 P_{c 2}\right)^m\right)} \\
P_{c 2}=\left(1-\tau_{s c f}\right)^{N_{s c f}}\left[1-\left(1-\tau_{m h}\right)^{N-N_{s c f}-1}\right]
\end{array}\right.
\end{equation}

The values of $\tau_{s c f}$, $\tau_{m h}$, $P_{c 1}$ and $P_{c 2}$ can be solved by a binary linear equation group combining \eqref{eq 20_1} and \eqref{eq 20_2}.

Let $P_{tr}$ be the probability that there is at least one transmission in the considered slot time, given by
\begin{equation}
\label{eq 21}
P_{t r}=1-\left(1-\tau_{s c f}\right)^{N_{s c f}}\left(1-\tau_{m h}\right)^{N-N_{s c f}}.
\end{equation}

With at least one UAV accessing the control channel, the probability of a transmission occurring successfully once on the control channel is $P_{s}$, which is given by
\begin{equation}
\label{eq 22}
\begin{aligned}
P_s= & \frac{N_{s c f} \tau_{s c f}\left(1-\tau_{s c f}\right)^{N_{s c f}-1}}{P_{t r}}+ \\
& \frac{\left(N-N_{s c f}\right) \tau_{m h}\left(1-\tau_{s c f}\right)^{N_{s f f}}\left(1-\tau_{m h}\right)^{N-N_{s c f}-1}}{P_{t r}}
\end{aligned}
.
\end{equation}

The network throughput can be expressed by $P_{tr}$ and $P_{s}$ as
\begin{equation}
\label{eq 23}
S=\frac{P_{s} P_{t r} E[P]}{\left(1-P_{t r}\right) \sigma+P_{s} P_{t r} T_{s}+P_{t r}\left(1-P_{s}\right) T_{c}},
\end{equation}
where $\sigma$ is the duration of an empty slot time, $E[P]$ is the average packet payload size, $T_s$ is the average time the channel is sensed busy because of a successful transmission, and $T_c$ is the average time the channel is sensed busy by each UAV during a collision. We obtain
\begin{equation}
\left\{\begin{array}{l}
T_{s}=R T S+S I F S+C T S+S I F S+C R T S+D I F S \\
T_{c}=R T S+D I F S
\end{array}\right.
.
\end{equation}

Considering the case of multiple data channels, there is an upper bound on the throughput when all data channels are occupied. The throughput will reach the upper limit when the node finds no resources for the data channel. The reason is that the packets sent are so long that the node occupies the data channel for a long time. To simplify the calculation, we assume that the length of the packets transmitted by UAVs satisfies 
\begin{equation}
\label{eq 25}
E[P] \leq T_{s} M r_{tr},
\end{equation}
where $M$ is the total number of data channels and $r_{tr}$ is transmission rate. With this constraint, we can ignore the impact of multiple data channels on throughput.

\section{Numerical Results}
\label{sim}
In this section, we first verify the fit between the theoretical and simulated values of the probability of being in SCF mode by 100,000 random scatters and observe the effect of $t$ and $d$ on the probability of being in SCF mode. Next, we simulate the theoretical and simulated values of the throughput of UD-MAC. Since VeMAC also adopts the idea of node classification access, we compare the performance of UD-MAC and VeMAC for analysis. Finally, we weigh the amount of data transmitted in SCF mode and MH mode.

We consider networks in which micro UAVs are used to perform tasks and $R = 5km$ \cite{sim1}. The wireless communication of the inter-UAV/UAV-GU uses IEEE 802.11g and other parameters are set as follows: $r = 100m$, $\bar{v} = 18km/h$, $M = 13$, $r_{tr} = 36Mbps$, RTS = $4.5\mu s$, CTS = $3.2 \mu s$, CRTS = $3.2 \mu s$, SIFS = $10 \mu s$, DIFS = $28 \mu s$, $\sigma = 9\mu s$, E[P] = 27$kbit$, and $N = 100$. We first scatter 100,000 points in 1-D, 2-D, and 3-D to simulate the location of the returning UAV at any given moment. Then the probability of being in SCF mode is obtained by the ratio of the number of points within the waiting range of the non-returning UAV to the total number of points. When simulating the throughput, we first divide the total nodes into the number of nodes in SCF mode and the number of nodes in MH mode based on the probability of being in SCF mode. Then we simulate the competition process of nodes and counted the throughput. The number of nodes and the competition window are large enough to ensure that the probability of one conflict is independent and unique \cite{throughout_1}, which makes the simulation close to the theory.

\begin{figure}[t]
	\centering
	\includegraphics[width=0.3\textheight]{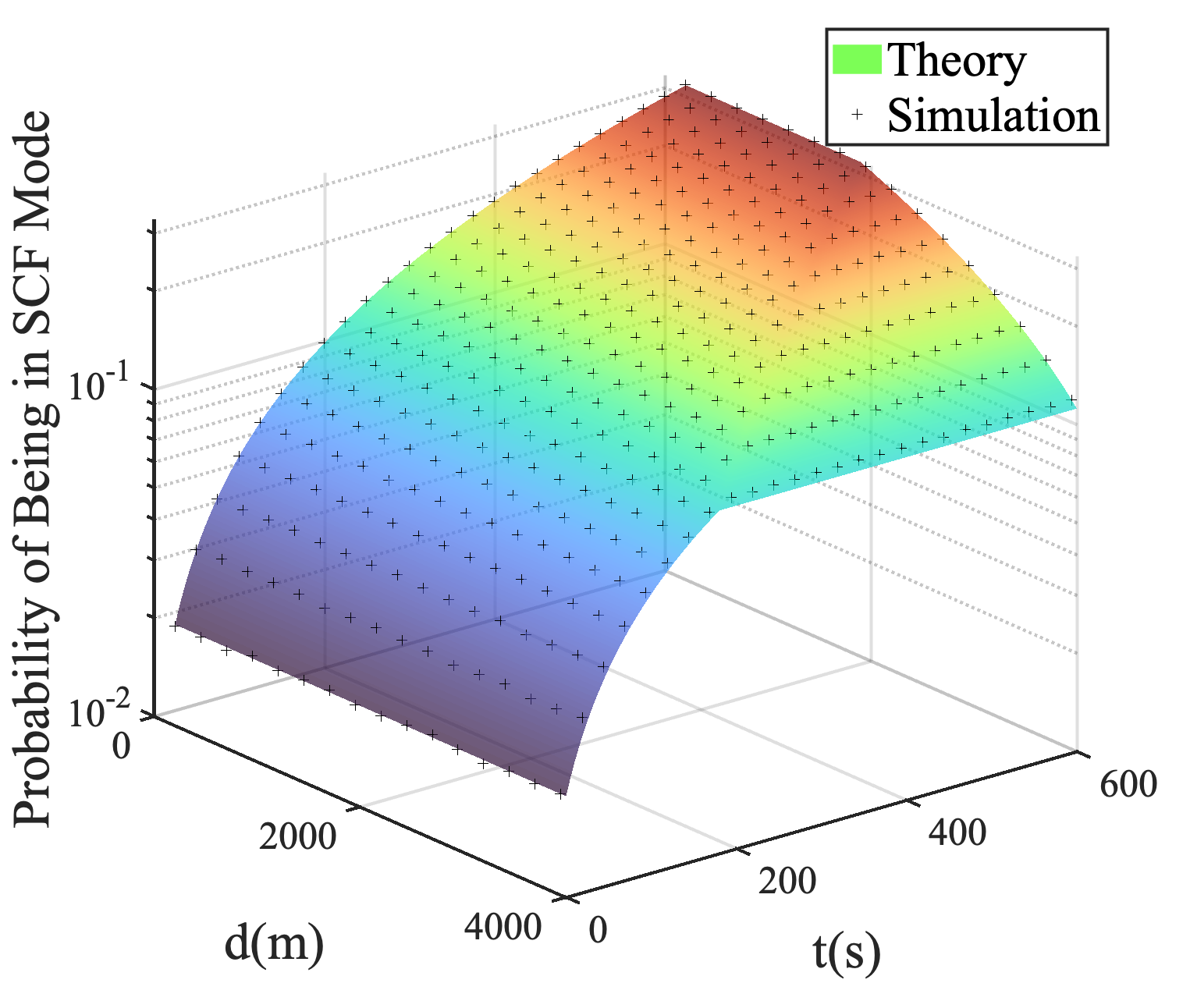}
	\DeclareGraphicsExtensions.
	\caption{The Probability of Being in SCF Mode in 1-D.}
	\label{fig 12}
\end{figure}

\begin{figure}[t]
	\centering
	\includegraphics[width=0.28\textheight]{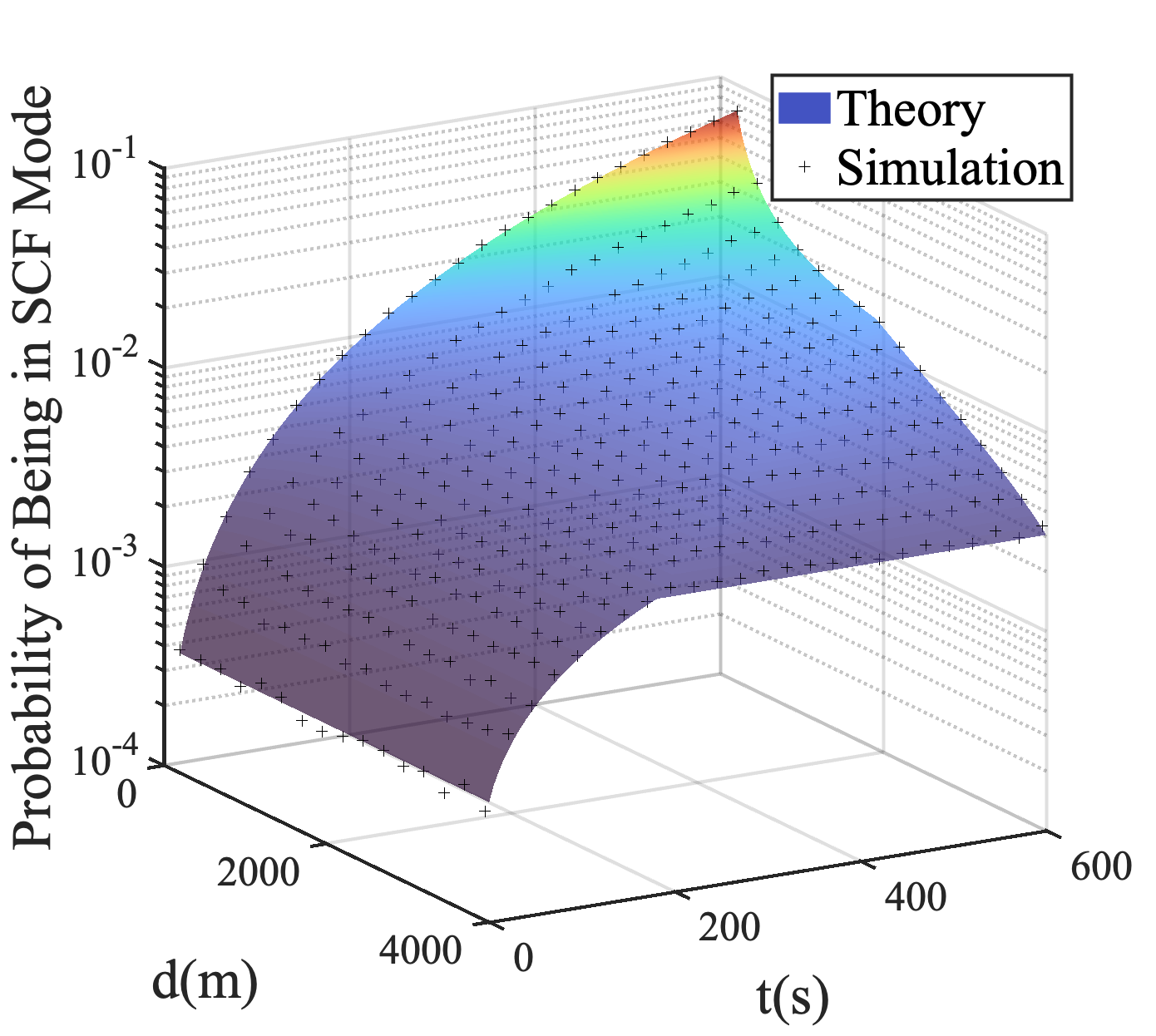}
	\DeclareGraphicsExtensions.
	\caption{The Probability of Being in SCF Mode in 2-D.}
	\label{fig 13}
\end{figure}

\begin{figure}[t]
	\centering
	\includegraphics[width=0.28\textheight]{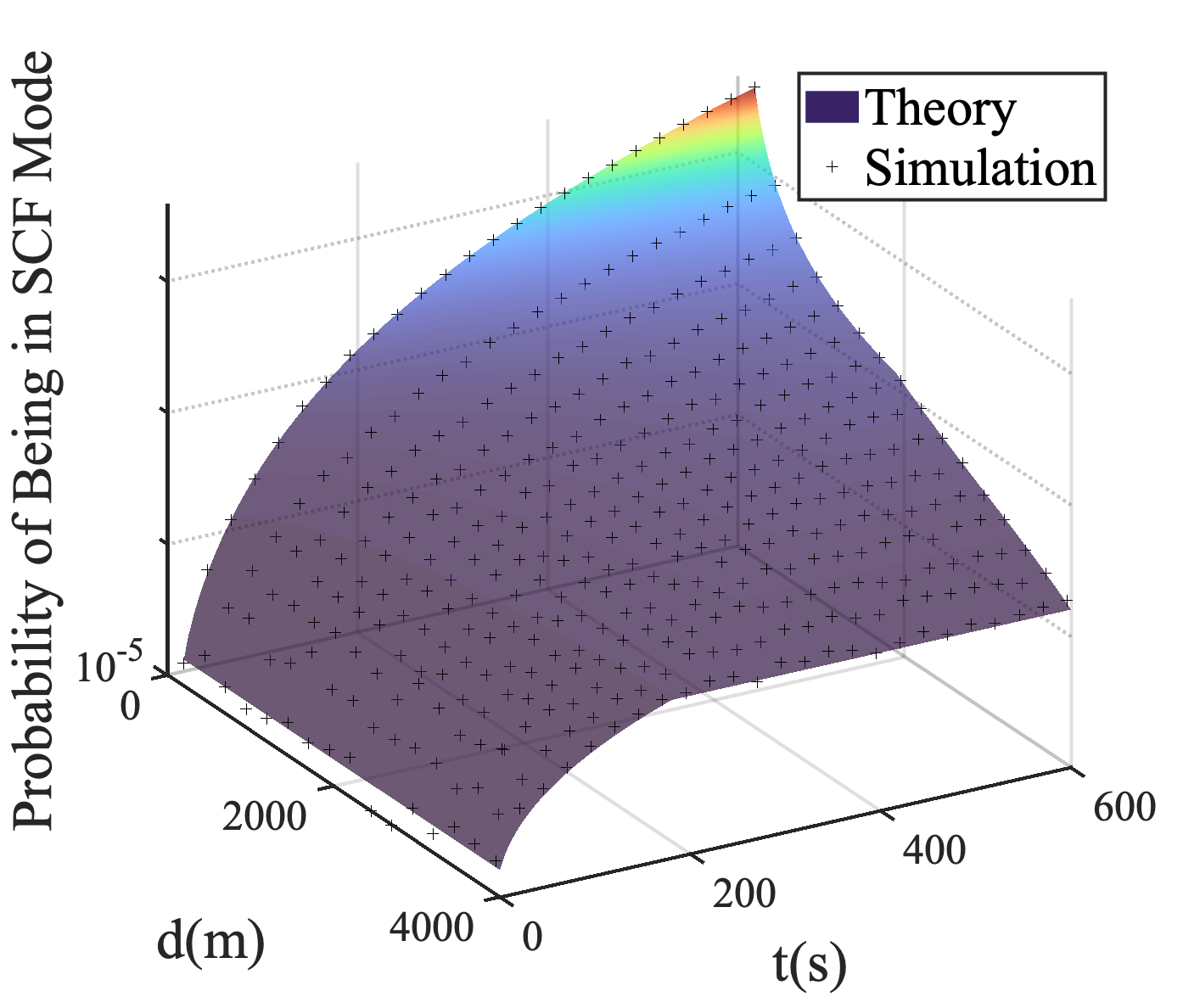}
	\DeclareGraphicsExtensions.
	\caption{The Probability of Being in SCF Mode in 3-D.}
	\label{fig 14}
\end{figure}
As shown in \textcolor[rgb]{0,0,0}{Fig. \ref{fig 12}}, \textcolor[rgb]{0,0,0}{Fig. \ref{fig 13}}, and \textcolor[rgb]{0,0,0}{Fig. \ref{fig 14}}, to facilitate observation, we use 3-D surfaces to represent theoretical values for the probability of being in SCF mode and points to represent the simulated values. The maximum errors between the theoretical and simulated values are about 0.84\% (in 1-D), 1.58\% (in 2-D), and 1.31\% (in 3-D). We prove that the theoretical values and the simulation values fit well. We find that the probability of a single UAV being in SCF mode is the highest when UAVs' space of activity is 1-D, and the probability is mostly greater than $0.1$. The probability of a single UAV being in SCF mode in 3-D is the lowest, and the probability is mostly close to $0$. This is because the gap between $r$ and $R$ is magnified in 3-D. We consider $r/R=0.02$ in 1-D, and $(r/R)^{3}=8\times10^{-6}$ in 3-D, which can be seen as reducing the probability of being in SCF mode in 3-D by $(r/R)^{2}=4\times10^{-4}$ times. It should be noted that we only provide the probability that a single UAV is in SCF mode under the uniform distribution, and it does not represent the actual probability of being in SCF mode. In addition, the closer the UAV is to GU, the longer the waiting time $t$, and the greater the probability of being in SCF mode. It is evident that the probability of being in SCF mode does not increase with $t$ when $d \in (3000,4000)$. This is because the condition that the non-returning UAV cannot appear outside the space of the scene limits the increase in the probability of being in SCF mode. In practice, the size of the wait time $t$ can be controlled by us, while the distance $d$ is not. Fortunately, $t$ affects the probability of being in SCF mode more than $d$. Therefore, we will do subsequent simulations by changing $t$.

\begin{figure}[t]
	\centering
	\includegraphics[width=0.31\textheight]{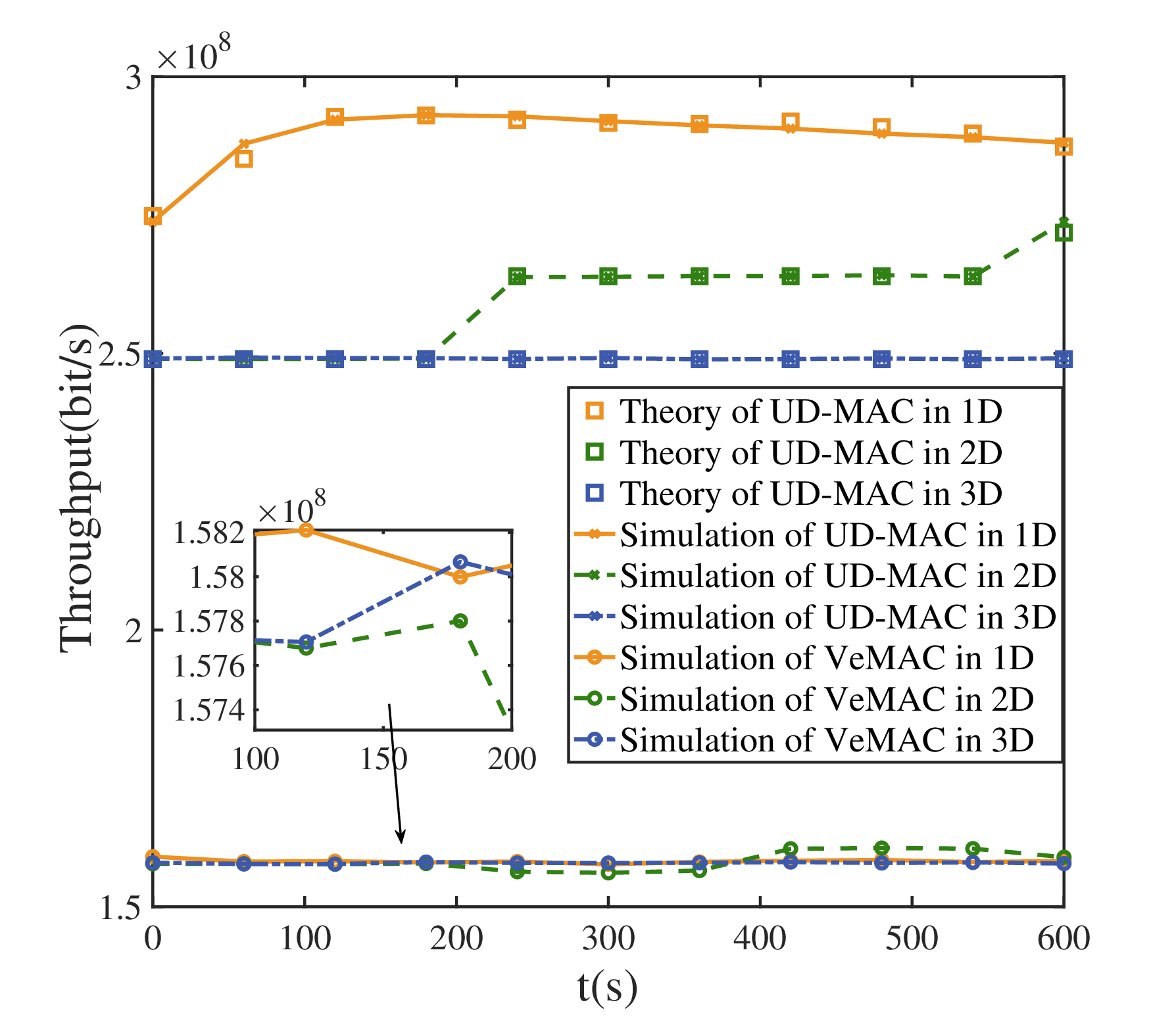}
	\DeclareGraphicsExtensions.
	\caption{Comparison of Throughput Between UD-MAC and VeMAC.}
	\label{fig 15}
\end{figure}

VeMAC divides the vehicles into two sets according to their direction of motion \cite{rela_1}. Vehicles in different sets can access the time slots of different sets. According to the VeMAC protocol, time slots are divided into two parts. A part is accessed by UAVs in SCF mode, and another part is accessed by UAVs in MH mode. The ratio of the size of the two parts is equal to the ratio of the number of UAVs in SCF mode to the number of UAVs in MH mode. The simulation in \textcolor[rgb]{0,0,0}{Fig. \ref{fig 15}} compares the throughput of UD-MAC and VeMAC, and indicates that the throughput of UD-MAC can be improved by 57\% to 83\% compared to VeMAC. Therefore, UD-MAC suits networks with high-capacity transmission requirements more than VeMAC. As shown in \textcolor[rgb]{0,0,0}{Fig. \ref{fig 15}}, the throughput curves of VeMAC fluctuate slightly in 1-D, 2-D, and 3-D. When the probability of being in SCF mode is relatively low, the idea of VeMAC's slot division cannot significantly improve network throughput. This is because the number of UAVs in SCF mode is far less than the total number of UAVs, and the slot division method degenerates into a traditional slot random access method.

Next, we observe the simulation curves of UD-MAC throughput in 1-D, 2-D, and 3-D. Because the maximum error between the theoretical and simulated values is about 0.94\%, the theoretical and simulation results of the throughput of UD-MAC fit well. The probability of a single UAV being in SCF mode in 3-D is very low, so that the number of UAVs in SCF mode within two hops is 0. Hence, there is no significant gain in network throughput in 3-D. However, when the gap between the activity area and the communication range becomes smaller, the throughput gain of SCF mode in 3-D will be more obvious, like in 1-D and 2-D scenarios. Since the number of UAVs in SCF mode within two hops is a positive integer, the resulting curve in 2-D is an ascending step curve. Whether the curve will continue to rise in the future can be concluded from the resulting curve in 3-D. Due to the high probability of being in SCF mode in 1-D, the number of UAVs in SCF mode within two hops increases as the probability of being in SCF mode increases. It is obvious that throughput increases first and then decreases as the probability of being in SCF mode increases. This is because when the number of UAVs in SCF mode is small, collisions between UAVs in SCF mode rarely occur. As the number of UAVs in SCF mode increases, the probability of collisions between UAVs in SCF mode increases, resulting in a decrease in network throughput. But the decline in throughput is gradual. When the number of UAVs in SCF mode equals the total number of UAVs, UD-MAC will degenerate into traditional CSMA/CA. The curve will descend until it coincides with the throughput simulation curve in 3-D. Thus, we can utilize the mobility of UAVs to achieve significant throughput improvements compared to traditional CSMA/CA.

\begin{figure}[t]
	\centering
	\includegraphics[width=0.32\textheight]{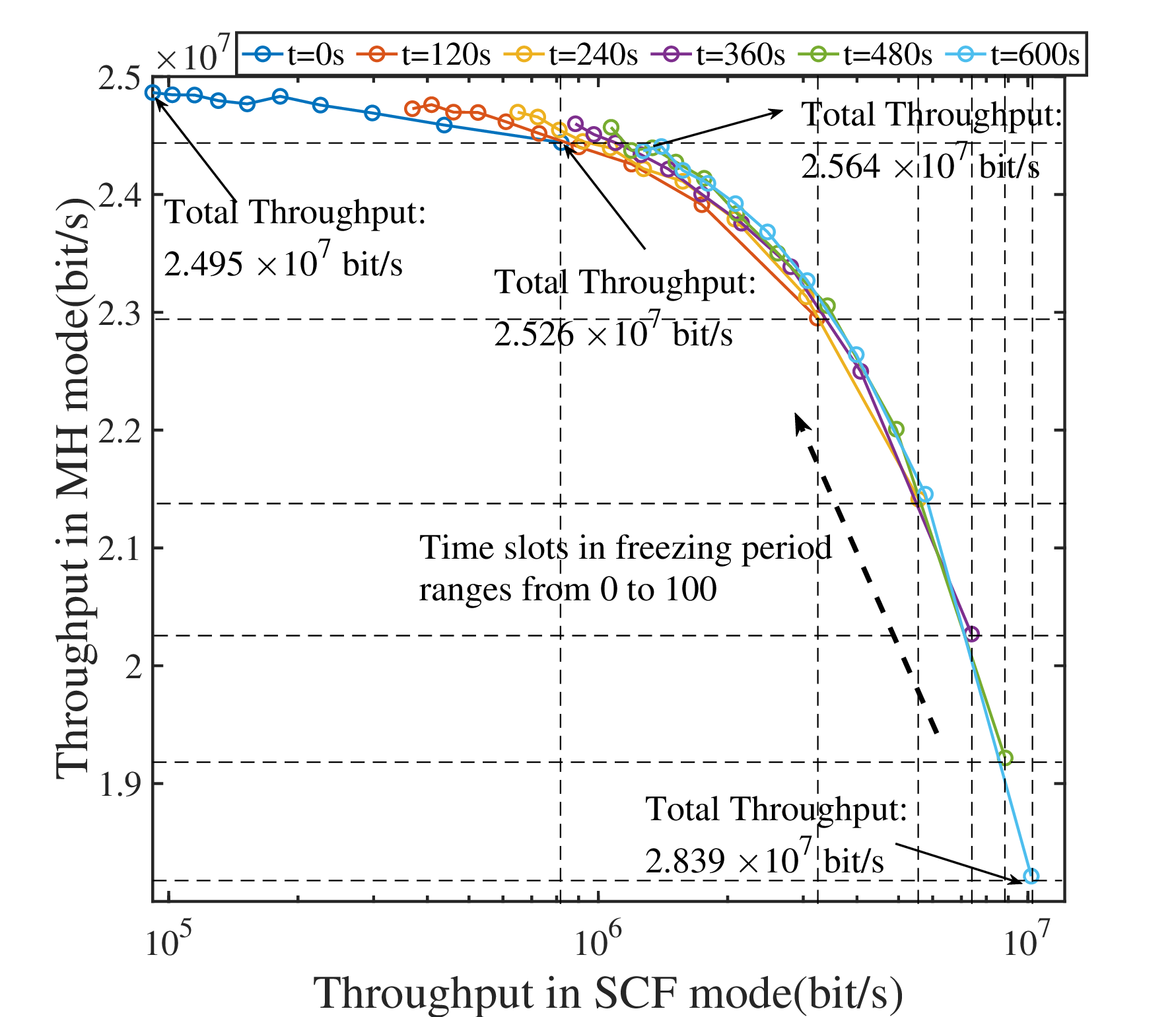}
	\DeclareGraphicsExtensions.
	\caption{The Tradeoff Between Throughput in SCF Mode and in MH Mode.}
	\label{fig 16}
\end{figure}

 As shown in \textcolor[rgb]{0,0,0}{Fig. \ref{fig 16}}, we simulate the throughput change curves of SCF mode and MH mode when $t$ is different. With the increase of $t$, UAVs in SCF mode will consume more network resources, increasing the total throughput. Since the MH mode can transmit data to the GU for nodes without SCF opportunities, the SCF mode cannot always occupy network resources. Therefore, we change the number of time slots during a freezing period based on the throughput in SCF mode and MH mode.  
When the number of time slots in the freezing period is 100, the data of the six curves are concentrated on the top in \textcolor[rgb]{0,0,0}{Fig. \ref{fig 16}}. It shows that we can allocate more network resources to UAVs in MH mode by increasing the number of time slots in the freezing period.  We find that increasing the number of time slots during the freezing period is detrimental to the total throughput. The effect of the freezing period on total throughput is more obvious when $t$ is large. Waiting time and freezing period are a pair of opposite influence factors on the total throughput. Therefore, we can use the waiting time and the freezing period to cooperatively control the resources occupied by SCF mode and MH mode to meet the dynamic performance requirements of nodes.

\section{Conclusion}
In this paper, we use a Markov state model to analyze the throughput of UD-MAC. To provide a theoretical reference for the analysis of network throughput, we analyze the probability of being in SCF mode when the UAVs' space of activities is 1-D, 2-D, and 3-D. The simulation results demonstrate that UD-MAC improves network throughput by utilizing the mobility of UAVs to achieve a throughput of 57\% to 83\% higher than VeMAC. In addition, we find that we can balance the trade-off the throughput of SCF and MH mode by an adaptive freezing period. Using the SCF mode to forward packets to any receiver, not just GU, can achieve higher network throughput, which is the future work of UD-MAC.


\begin{thebibliography}{00}

\bibitem{intro_1}
P. Sun, A. Boukerche and Y. Tao, ``Theoretical Analysis of the Area Coverage in a UAV-based Wireless Sensor Network," 2017 13th International Conference on Distributed Computing in Sensor Systems (DCOSS), 2017, pp. 117-120, doi: 10.1109/DCOSS.2017.18.
\bibitem{intro_2}
B. T. Fraser and R. G. Congalton, ``Issues in unmanned aerial systems (uas) data collection of complex forest environments", Remote Sensing, vol. 10, no. 6, pp. 908, 2018.
\bibitem{intro_3}
J. R. Kellner, J. Armston, M. Birrer, K. Cushman, L. Duncanson, C. Eck, C. Falleger, B. Imbach, K. Král, M. Krček et al., ``New opportunities for forest remote sensing through ultra-high-density drone lidar", Surveys in Geophysics, vol. 40, no. 4, pp. 959-977, 2019.
\bibitem{intro_4}
K. Joyce, S. Duce, S. Leahy, J. Leon and S. Maier, ``Principles and practice of acquiring drone-based image data in marine environments", Marine and Freshwater Research, vol. 70, no. 7, pp. 952-963, 2019.
\bibitem{intro_5}
H. Kuang, H. Cao, X. Li and H. Cheng, ``A Framework for Multi-Event Data Collection Using Unmanned Aerial Vehicle Aided Internet of Things in Smart Agriculture," 2021 IEEE 2nd International Conference on Information Technology, Big Data and Artificial Intelligence (ICIBA), 2021, pp. 174-177, doi: 10.1109/ICIBA52610.2021.9688027.
\bibitem{new_7}
C. Fan, X. Zhou, T. Zhang, W. Yi and Y. Liu, ``Cache-Enabled UAV Emergency Communication Networks: Performance Analysis With Stochastic Geometry," in IEEE Transactions on Vehicular Technology, doi: 10.1109/TVT.2023.3249283.
\bibitem{new_1}
Cambra, Carlos, et al. ``Ad hoc network for emergency rescue system based on unmanned aerial vehicles." Network Protocols and Algorithms 7.4 (2015): 72-89.
\bibitem{intro_6}
M. Rossi and D. Brunelli, ``Autonomous gas detection and mapping with unmanned aerial vehicles", IEEE Trans. Instrum. Meas., vol. 65, no. 4, pp. 765-775, Apr. 2016.
\bibitem{intro_7}
X. Li, J. Tan, A. Liu, P. Vijayakumar, N. Kumar and M. Alazab, ``A Novel UAV-Enabled Data Collection Scheme for Intelligent Transportation System Through UAV Speed Control," in IEEE Transactions on Intelligent Transportation Systems, vol. 22, no. 4, pp. 2100-2110, April 2021, doi: 10.1109/TITS.2020.3040557.
\bibitem{new_5}
Y. Xu, T. Zhang, Y. Liu and D. Yang, ``UAV-Enabled Integrated Sensing, Computing, and Communication: A Fundamental Trade-Off," in IEEE Wireless Communications Letters, vol. 12, no. 5, pp. 843-847, May 2023, doi: 10.1109/LWC.2023.3245728.
\bibitem{intro_71}
Y. Qin, W. Yang and L. Peng, ``Inter-Session Network Coding with Clustering Routing in Wireless Delay Tolerant Networks," 2019 20th Asia-Pacific Network Operations and Management Symposium (APNOMS), Matsue, Japan, 2019, pp. 1-4, doi: 10.23919/APNOMS.2019.8893089.

\bibitem{intro_8}
Y. Zeng and R. Zhang, ``Energy-Efficient UAV Communication With Trajectory Optimization," in IEEE Transactions on Wireless Communications, vol. 16, no. 6, pp. 3747-3760, June 2017, doi: 10.1109/TWC.2017.2688328.
\bibitem{intro_9}
Y. Zeng, R. Zhang and T. J. Lim, ``Throughput maximization for UAV-enabled mobile relaying systems", IEEE Trans. Commun., vol. 64, no. 12, pp. 4983-4996, Dec. 2016.

\bibitem{new_3}
Cheng, Fen, et al. ``UAV trajectory optimization for data offloading at the edge of multiple cells." IEEE Transactions on Vehicular Technology 67.7 (2018): 6732-6736.
\bibitem{new_4}
Liu, Xiaonan, et al. "Placement and power allocation for NOMA-UAV networks." IEEE Wireless Communications Letters 8.3 (2019): 965-968.
\bibitem{new_6}
Y. Xu, T. Zhang, D. Yang, Y. Liu and M. Tao, ``Joint Resource and Trajectory Optimization for Security in UAV-Assisted MEC Systems," in IEEE Transactions on Communications, vol. 69, no. 1, pp. 573-588, Jan. 2021, doi: 10.1109/TCOMM.2020.3025910.
\bibitem{intro_10}
Q. Wu, L. Liu and R. Zhang, ``Fundamental Trade-offs in Communication and Trajectory Design for UAV-Enabled Wireless Network," in IEEE Wireless Communications, vol. 26, no. 1, pp. 36-44, February 2019, doi: 10.1109/MWC.2018.1800221.
\bibitem{intro_11}
M. Y. Arafat and S. Moh, ``Location-Aided Delay Tolerant Routing Protocol in UAV Networks for Post-Disaster Operation," in IEEE Access, vol. 6, pp. 59891-59906, 2018, doi: 10.1109/ACCESS.2018.2875739.
\bibitem{intro_12}
X. Tan, Z. Zuo, S. Su, X. Guo, X. Sun and D. Jiang, ``Performance Analysis of Routing Protocols for UAV Communication Networks," in IEEE Access, vol. 8, pp. 92212-92224, 2020, doi: 10.1109/ACCESS.2020.2995040.
\bibitem{intro_13}
J. Baek, S. I. Han and Y. Han, ``Energy-Efficient UAV Routing for Wireless Sensor Networks," in IEEE Transactions on Vehicular Technology, vol. 69, no. 2, pp. 1741-1750, Feb. 2020, doi: 10.1109/TVT.2019.2959808.
\bibitem{new_2}
Vashisht, Sahil, Sushma Jain, and Gagangeet Singh Aujla. ``MAC protocols for unmanned aerial vehicle ecosystems: Review and challenges." Computer Communications 160 (2020): 443-463.
\bibitem{intro_14}
Y. Zeng, R. Zhang and T. J. Lim, ``Wireless communications with unmanned aerial vehicles: opportunities and challenges," in IEEE Communications Magazine, vol. 54, no. 5, pp. 36-42, May 2016, doi: 10.1109/MCOM.2016.7470933.

\bibitem{intro_72}
Zhang, Zhensheng. ``Routing in intermittently connected mobile ad hoc networks and delay tolerant networks: overview and challenges." IEEE Communications Surveys \& Tutorials 8.1 (2006): 24-37.

\bibitem{UD-MAC}
X. Liu, Z. Wei, Z. Feng and F. Ning, ``UD-MAC: Delay tolerant multiple access control protocol for unmanned aerial vehicle networks," 2017 IEEE 28th Annual International Symposium on Personal, Indoor, and Mobile Radio Communications (PIMRC), Montreal, QC, Canada, 2017, pp. 1-6, doi: 10.1109/PIMRC.2017.8292602.


\bibitem{rela_01}
``IEEE Standard for Information technology--Telecommunications and information exchange between systems Local and metropolitan area networks--Specific requirements Part 11: Wireless LAN Medium Access Control (MAC) and Physical Layer (PHY) Specifications," in IEEE Std 802.11-2012 (Revision of IEEE Std 802.11-2007) , vol., no., pp.1-2793, 29 March 2012, doi: 10.1109/IEEESTD.2012.6178212.
\bibitem{rela_02}
IEEE Computer Society LAN MAN Standard Committee. ``Wireless LAN medium access control (MAC) and physical layer (PHY) specifications." IEEE Std. 802.11-1997 (1997).






\bibitem{rela_1}
H. A. Omar, Weihua Zhuang and Li Li, ``VeMAC: A novel multichannel MAC protocol for vehicular ad hoc networks," 2011 IEEE Conference on Computer Communications Workshops (INFOCOM WKSHPS), Shanghai, China, 2011, pp. 413-418, doi: 10.1109/INFCOMW.2011.5928848.
\bibitem{rela_2}
B. Li, F. Hou, C. Zhang, S. Li, S. Ji and S. Chen, ``MAC-AC: A Novel Distributed MAC Protocol for Accessing Channel in Vehicular Ad Hoc Networks," 2020 IEEE 92nd Vehicular Technology Conference (VTC2020-Fall), Victoria, BC, Canada, 2020, pp. 1-5, doi: 10.1109/VTC2020-Fall49728.2020.9348700.
\bibitem{rela_3}
T. Zhang and Q. Zhu, ``EVC-TDMA: An enhanced TDMA based cooperative MAC protocol for vehicular networks," in Journal of Communications and Networks, vol. 22, no. 4, pp. 316-325, Aug. 2020, doi: 10.1109/JCN.2020.000021.
\bibitem{rela_4}
B. Li, X. Guo, R. Zhang, X. Du and M. Guizani, ``Performance Analysis and Optimization for the MAC Protocol in UAV-Based IoT Network," in IEEE Transactions on Vehicular Technology, vol. 69, no. 8, pp. 8925-8937, Aug. 2020, doi:10.1109/TVT.2020.2997782.
\bibitem{rela_5}
H. Baek and J. Lim, ``Time Mirroring Based CSMA/CA for Improving Performance of UAV-Relay Network System," in IEEE Systems Journal, vol. 13, no. 4, pp. 4478-4481, Dec. 2019, doi: 10.1109/JSYST.2019.2912385.
\bibitem{rela_6}
Y. Kwon, H. Baek and J. Lim, ``Uplink NOMA Using Power Allocation for UAV-Aided CSMA/CA Networks," in IEEE Systems Journal, vol. 15, no. 2, pp. 2378-2381, June 2021, doi: 10.1109/JSYST.2020.3028884.

\bibitem{scf_1}
Y. Wang, Y. Liu, J. Zhang, H. Ye and Z. Tan, ``Cooperative Store–Carry–Forward Scheme for Intermittently Connected Vehicular Networks," in IEEE Transactions on Vehicular Technology, vol. 66, no. 1, pp. 777-784, Jan. 2017, doi: 10.1109/TVT.2016.2536059.
\bibitem{scf_2}
A. A. Siddig, A. S. Ibrahim and M. H. Ismail, ``Full-Duplex Store-Carry-Forward scheme for Intermittently Connected Vehicular Networks," 2020 IEEE 91st Vehicular Technology Conference (VTC2020-Spring), Antwerp, Belgium, 2020, pp. 1-6, doi: 10.1109/VTC2020-Spring48590.2020.9129598.
\bibitem{scf_3}
F. Nordemann and R. Tönjes, ``Transparent and autonomous store-carry-forward communication in Delay Tolerant Networks (DTNs)," 2012 International Conference on Computing, Networking and Communications (ICNC), Maui, HI, USA, 2012, pp. 761-765, doi: 10.1109/ICCNC.2012.6167525.
\bibitem{scf_4}
P. Kolios, V. Friderikos and K. Papadaki, ``Energy-Efficient Relaying via Store-Carry and Forward within the Cell," in IEEE Transactions on Mobile Computing, vol. 13, no. 1, pp. 202-215, Jan. 2014, doi: 10.1109/TMC.2012.233.
\bibitem{scf_5}
F. Xiong et al., ``Energy-Saving Data Aggregation for Multi-UAV System," in IEEE Transactions on Vehicular Technology, vol. 69, no. 8, pp. 9002-9016, Aug. 2020, doi: 10.1109/TVT.2020.2999374.

\bibitem{scf_6}
Z. Wei, H. Wu, S. Huang and Z. Feng, ``Scaling Laws of Unmanned Aerial Vehicle Network with Mobility Pattern Information," in IEEE Communications Letters, vol. 21, no. 6, pp. 1389-1392, June 2017, doi: 10.1109/LCOMM.2017.2671861.
\bibitem{scf_7}
H. Zheng, X. Li and N. Ye, ``Random Subchannel Selection of Store-Carry and Forward Transmissions in Traffic Hotspots," in IEEE Communications Letters, vol. 21, no. 9, pp. 2073-2076, Sept. 2017, doi:10.1109/LCOMM.2017.2705645.




\bibitem{throughout_1}
Bianchi, Giuseppe. "Performance analysis of the IEEE 802.11 distributed coordination function." IEEE Journal on selected areas in communications 18.3 (2000): 535-547.
\bibitem{sim1}
Gupta, Suraj G., Dr Ghonge, and Pradip M. Jawandhiya. ``Review of unmanned aircraft system (UAS)." International Journal of Advanced Research in Computer Engineering \& Technology (IJARCET) Volume 2 (2013).
\bibitem{sim2}
Using UAV Networks, ``2019 IEEE SmartWorld, Ubiquitous Intelligence \& Computing, Advanced \& Trusted Computing, Scalable Computing \& Communications, Cloud \& Big Data Computing, Internet of People and Smart City Innovation (SmartWorld/SCALCOM/UIC/ATC/CBDCom/IOP/SCI), Leicester, UK, 2019, pp. 1595-1600, doi: 10.1109/SmartWorld-UIC-ATC-SCALCOM-IOP-SCI.2019.00285.




\end{thebibliography}
\end{document}